\documentclass[pra,twocolumn,floatfix,nopacs,superscriptaddress]{revtex4-1}
\usepackage{graphicx,amsfonts,amssymb,amsmath}
\usepackage[displaymath,mathlines]{lineno}
\usepackage[dvipsnames]{xcolor}
\usepackage{here}
\usepackage{graphicx}
\definecolor{darkblue}{rgb}{0.0, 0.0, 0.75}
\usepackage{color}
\usepackage[colorlinks=true,
linkcolor=darkblue,
urlcolor=darkblue,
citecolor=darkblue]{hyperref}

\newcommand{\comment}[1]{}

\newcommand*{\bra}[1]{\mathopen{\langle}#1\mathclose{|}}
\newcommand*{\ket}[1]{\mathopen{|}#1\mathclose{\rangle}}

\newcommand{\ketbrap}[2]{\ket{#1}\hspace{-0.25em}\bra{#2}}

\begin{document}
	
\title{Efficient computation of topological order}
	
\author{Louis Fraatz}
\email{louis.fraatz@tu-berlin.de}
\affiliation{Institut f\"ur Physik und Astronomie, Technische Universit\"at Berlin, Hardenbergstra{\ss}e 36 EW 7-1, 10623 Berlin, Germany}
\author{Amit Jamadagni}
\affiliation{Quantum Computing and Technologies Department, Leibniz Supercomputing Centre (LRZ), 85748 Garching, Germany}
\author{Hendrik Weimer}
\affiliation{Institut f\"ur Physik und Astronomie, Technische Universit\"at Berlin, Hardenbergstra{\ss}e 36 EW 7-1, 10623 Berlin, Germany}

\begin{abstract}

  We analyze the computational aspects of detecting topological order
  in a quantum many-body system. We contrast the widely used
  topological entanglement entropy with a recently introduced
  operational definition for topological order based on
  error correction properties, finding exponential scaling with the
  system size for the former and polynomial scaling for the latter. We
  exemplify our approach for a variant of the paradigmatic toric code
  model with mobile particles, finding that the error correction
  method allows to treat substantially larger system sizes. In
  particular, the phase diagram of the model can be successfully
  computed using error correction, while the topological entanglement
  entropy is too severely limited by finite size effects to obtain
  conclusive results. While we mainly focus on one-dimensional systems
  whose ground states can be expressed in terms of matrix product
  states, our strategy can be readily generalized to higher dimensions
  and systems out of equilibrium, even allowing for an efficient
  detection of topological order in current quantum simulation
  experiments.

\end{abstract}
	
\maketitle
	

Topologically ordered states of matter transcend Landau's
symmetry-breaking paradigm and cannot be described by local order
parameters \cite{Wen2019}. While such states are both important from a
fundamental point of view and for applications in quantum computing,
computing the topological properties of a many-body state is a
notoriously difficult task. Here, we show that an approach based on
the error correction properties of a state allows for an efficient
computation of the phase boundaries of topologically ordered states,
requiring only polynomial computational time in the size of the
system.

While there are many conceptual frameworks to identify topological
order
\cite{Thouless1982,Kohmoto1985,denNijs1989,Haegeman2012,Pollmann2012,Elben2020,Chen2010,Preskill2006,Levin2006,Balents2012,Zhang2012,Zhu2013,Nussinov2009,Qiu2020},
their use in actual calculations or even experiments remains
limited. The most pratically relevant quantity is the topological
entanglement entropy
\cite{Preskill2006,Levin2006,Balents2012,Zhang2012}, but being a
highly nonlinear functional of a reduced density matrix, it can only
be evaluated for small cluster sizes \cite{Satzinger2021}. This
underlines that the computational decision whether a state is
topologically ordered or not remains an outstanding problem.

In this Letter, we show that these challenges can be overcome when
computing topological properties based on an operational definition of
topological order related to the error correction properties of a
many-body state \cite{Jamadagni2022}. We exemplify this approach for a
variant of the toric code, where the consituent spins can move around
on a lattice. Using matrix product state (MPS) simulations, we show
that the ground state undergoes  two distinct phase
transitions. We emphasize that this behavior cannot reliably captured
using the topological entanglement entropy, which shows a continuous
dependence on the strength of the hopping of the particles rather than
converging to a discrete value.

\emph{Topological Order.---} Quantum phase transitions involving local order parameters can be successfully classified within Landau's symmetry breaking paradigm \cite{Sachdev2011}. However, in the last decades, many examples have been discovered that transcend this notion. Most prominently, gapped quantum phases exhibiting topological order \cite{Wen2017} exhibit a wide range of unrivaled properties such as long-range entanglement, robustness against local perturbations and the ability to perform fault-tolerant quantum computing using such many-body states.

One major challenge surrounding topological order is that its inherent
nonlocality makes the classification of topologically ordered states
of matter very difficult. On the one hand, this refers to
\emph{conceptual} challenges, i.e., identifying which nonlocal
quantities should be considered, especially for systems approaching
the thermodynamic limit of infinite system sizes. On the other hand
lie \emph{practical} challenges concerning the formalization of a
decision problem allowing to compute whether a given many-body state
is topologically ordered or not. For the latter, it is of special
importance that the decision problem can be solved efficiently, i.e.,
computational resources scale at most polynomially in the system
size, as otherwise only very small systems can be probed.

\textit{Topological entanglement entropy and its li\-mi\-ta\-tions.---} While there are many different concepts to describe topological order \cite{Thouless1982,Kohmoto1985,denNijs1989,Haegeman2012,Pollmann2012,Elben2020,Zhang2012,Zhu2013}, resorting to the topological entanglement entropy (TEE) \cite{Kitaev2006,Levin2006} is one of the most widely used approaches. In its essence, the TEE $\gamma$ captures the subleading correction to the entanglement entropy when looking at the reduced density matrix of a system cut along a path of length $L$, i.e.,
\begin{equation}
  S = \alpha L - \gamma + O(1/L).
  \label{eq:tee}
\end{equation}
The TEE is connected to the total quantum dimension of a topologically
ordered state \cite{Kitaev2006} and hence directly captures
topological order. For practical calculations, the TEE can be computed
from a cylindrical geometry \cite{Balents2012}, see Fig.~1a. In many
cases, it is possible to derive the correct value of the
two-dimensional (2D) case in the limit where the $Z$ extension of the
cylinder is very small \cite{Zou2016}. One important requirement for
this to work is that the model does not include boundary terms that
immediately destroy topological order \cite{Jamadagni2018}, i.e., the
quasi-1D model is a variant of symmetry-protected topological order
\cite{Chen2011} corresponding to the intrinsically topologically
ordered 2D model.
\begin{figure}[t]
	\centering

	\includegraphics[width=\linewidth]{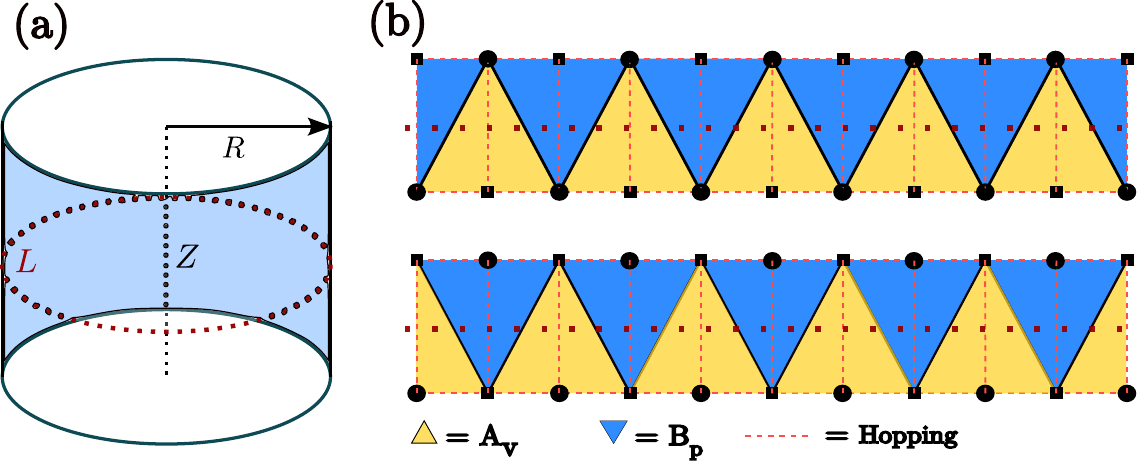}

	\caption{Sketch of the topology. A cylindric system (a) is
          mapped onto a quasi-1D ladder geometry in the limit of small
          $Z$ (b). The cut of length $L$ for computing the
          topological entanglement entropy is shown as a dotted
          line. For the mobile toric code, the system consists of two
          different sublattices (circles and squares) on which $A_v$
          and $B_p$ operators act to implement the toric code
          interactions. Mobility is introduced by allowing the
          particles to hop along the dashed lines.}
	\label{fig:zylinder}
\end{figure}

Unfortunately, actually obtaining the TEE is often prohibitively
challenging. In general, computing the entanglement entropy of a
many-body system scales exponentially in the system size. This is true
even in a quasi-1D situation, where the ground state can be
approximately represented by a matrix product state (MPS) of the form
\cite{Garcia2007}
\begin{equation}
\ket{\psi}=\sum_{i_1,....i_N}^d\text{Tr}[A_{i_1}^{[1]}A_{i_2}^{[2]}......A_{i_N}^{[N]}]\ket{i_1,i_2,....,i_N},
\end{equation}
where $\ket{\psi}$ is a state of the Hilbert space of $N$ particles of
local Hilbert space dimension $d$. The matrices $A_i$ have a maximum
size of $\chi \times \chi$, with $\chi$ being the maximum bond
dimension, which is a measure of complexity of the many-body
state. While for many 1D problems, $\chi$ grows at most polynomially
with system size, allowing to compute expectation values in polynomial
time, this is typically not the case for computing the TEE
\cite{Hamma2008b}. Generically, the cost of computing the entanglement
entropy scales exponentially in the number of bonds broken by the cut,
resulting in $O(d^{3N/2})$ steps for the cut required for the TEE on a
square ladder, see Fig.~1b. In practice, this means that it is only
possible to use the TEE for computing topological order for spin $1/2$
systems having $d=2$ and provided that the TEE converges sufficiently
fast in the system size.

\emph{Operational definition of topological order.---} To obtain a
computationally viable approach to topological order, we turn to a
recently introduced operational definition \cite{Jamadagni2022}. In a
nutshell, a state is topologically ordered if and only if it can be
corrected to a topologically ordered reference state using an error
correction circuit of finite depth. A reference state is state is
topologically ordered if it can be used to protect quantum information
against all local error operations. The circuit depth refers to the
number of steps that are required for the circuit to complete in a
fully parallelized implementation. In our approach, the reference
state takes a similar role as the order parameter in ordered phases
involving spontaneous symmetry breaking. We note that this operational
definition is related to efforts to enhance the topological order in
quantum states by means of error correction \cite{Cong2024}. The error
correction algorithm can be formulated as a sequence of (i) syndrome
measurements to detect the errors with respect to the reference state
and (ii) quantum operations to correct the errors. In sharp contrast
to the TEE, the operational definition does not require the
computation of a nonlinear functional and hence can be implmented in
polynomial time within an MPS framework. For practical purposes, the
most convenient implementation of the operational definition
corresponds to a Monte-Carlo algorithm \cite{Jamadagni2022a}: Starting
from the MPS state, $O(N)$ projective measurements are being performed
on the state, randomly selecting a measurement outcome according to
the respective expectation values. After each projection, the MPS has
to be reorthogonalized using $O(N)$ singular value decompositions
(SVDs) \cite{Orus2014}. Each SVD carries a computational cost of
$O(d\chi^3)$, leading to a total computational cost of
$O(dMN^2\chi^3)$, with $M$ being the number of Monte-Carlo
samples. This polynomial scaling demonstrates a drastic improvement
over the exponential complexity of computing the TEE.

\emph{The mobile toric code.---} Let us now turn to a concrete model
to apply our error correction method and compute its phase
diagram. Here, we will be interested in an extension to the toric
code, where the constituents particles can move around on a square
lattice, see Fig.~1b. As an experimental realization, we envision an
implementation of the toric code using ultracold atoms in an optical
lattice undergoing Rydberg interactions \cite{Weimer2010}, where
residual movement of the atoms is a natural imperfection stemming from
the finiteness of the optical lattice potentials \cite{Bloch2008}.
The toric code is described by the spin-$1/2$ Hamiltonian
\begin{equation}
H_\text{toric}=  -\sum_{\text{vertex}}\underset{A_v}{\underbrace{\sigma^i_x \sigma^j_x \sigma^k_x \sigma^l_x}} -\sum_{plaquette}\underset{B_p}{\underbrace{\sigma^m_z \sigma^n_z \sigma^o_z \sigma^p_z}}.
\end{equation}
Since all $A_v$ and $B_p$ operators mutually commute, the ground
states satisfy the constraints $A_v \ket{\psi}=\ket{\psi}$ and
$B_p\ket{\psi}$ for all $v$ and $p$ \cite{Kitaev2003}. Violations of
the constraints correspond to errors that can be detected by measuring
the $A_v$ and $B_p$ operators.

In order to account for the motion of particles, we extend the local
Hilbert space by a third possible state $\ket{0}$ indicating the
absence of a particle. In the following, we assume a hard-core
constraint for the particles, i.e., there is at most one atom per
lattice site, which is well justified in sufficiently deep optical
lattices \cite{Bloch2008}. Then, the local basis for each lattice site is given by the set $\{\ket{\downarrow}, \ket{0}, \ket{\uparrow}\}$, where $\ket{\uparrow}$ and $\ket{\downarrow}$ refer to an atom in the
spin up or down state, respectively. Using this notation, we can define bosonic creation and annhilation
operators according to
\begin{equation}
  a_\uparrow^\dagger =
  \ketbrap{\uparrow}{0},\;\;\;\;\;a_\downarrow^\dagger =
  \ketbrap{\downarrow}{0}.
\end{equation}
 The Pauli spin operators can then
be understood in the usual way, e.g., $\sigma_x =
\ketbrap{\uparrow}{\downarrow} + \text{H.c.}$, with the only
difference being that they are now acting on a three-dimensional local
Hilbert space. As a consequence, $A_v$ and $B_p$ operators now also
have eigenstates with eigenvalue of zero; this occurs whenever one of
the sites taking part in the vertex $v$ or the plaquette $p$ is in the
vacuum state $\ket{0}$. Then, the total Hamiltonian has the form
\begin{equation}
H= H_\text{toric}-t \sum_{\langle ij\rangle\;s}a_{i,s} a^{\dag}_{j,s} +\text{H.c.},
\end{equation}
where the sum over $s$ runs over both spin states. Since the square
lattice has twice as many lattice sites as the checkerboard lattice of
the original toric code, the toric code part has to be doubled to act
on both sublattices, see the top and bottom part of
Fig.~1b. In all cases, we work consider half filling
  of the lattice sites.

As discussed previously, we focus on a quasi-1D realization of the
model. Similar to the toric code in a magnetic field
\cite{Jamadagni2022}, the minimal instance of the mobile toric code is
given by a ladder geometry, where the four-body interactions of the
original model are replaced by three-body interactions, see
Fig.~1b. Before performing the error correction analysis, let us first
remark on some general properties of the phase diagram. For zero
hopping $t$, the system corresponds exactly to the quasi-1D limit of
the toric code and hence exhibits intrinsic topological order when
extended back into 2D, see App.~A. Additionally, the
doubling of the lattice sites creates an additional $\mathbb{Z}_2$
symmetry corresponding to which of the two sublattices are being
populated with atoms, with the other half being empty for $t=0$.  For
very large values of $t$, the toric code part irrelevant and the
ground state corresponds to a topologically trivial Luttinger liquid
phase. At intermediate $t$, both the topological order and the
density-wave order of the broken $\mathbb{Z}_2$ symmetry will
eventually disappear. However, it is possible that both types of order
disappear at the same time or there is an intermediate phase
exhibiting density-wave order but no topological order. Hence, we will
employ our computationally efficient strategy to determine the entire
phase diagram.

\emph{Error correction circuits.---} For the mobile toric code, the
error correction algorithm consists of two parts. First, the positions
of the particles have to be corrected to a perfect crystal before the
conventional error correction to the toric code can be
applied. Immediately applying the error correction for the toric code
would result in some $A_v$ and $B_p$ operators being measured in the
zero eigenvalue, which cannot be corrected. Importantly, the depth of
the error correction circuit for correcting the positions of the
particle is a measure of density-wave ordering
\cite{Jamadagni2022a}. If this part of the error correction circuit
has finite depth, the system exibits density-wave ordering; if the
circuit depth diverges with the size of the system and the system
enters a Luttinger liquid phase. Hence, density-wave ordering is a
necessary requirement for the system being able to exhibit topological
order.

\begin{figure}[t]
  \includegraphics[width=\linewidth]{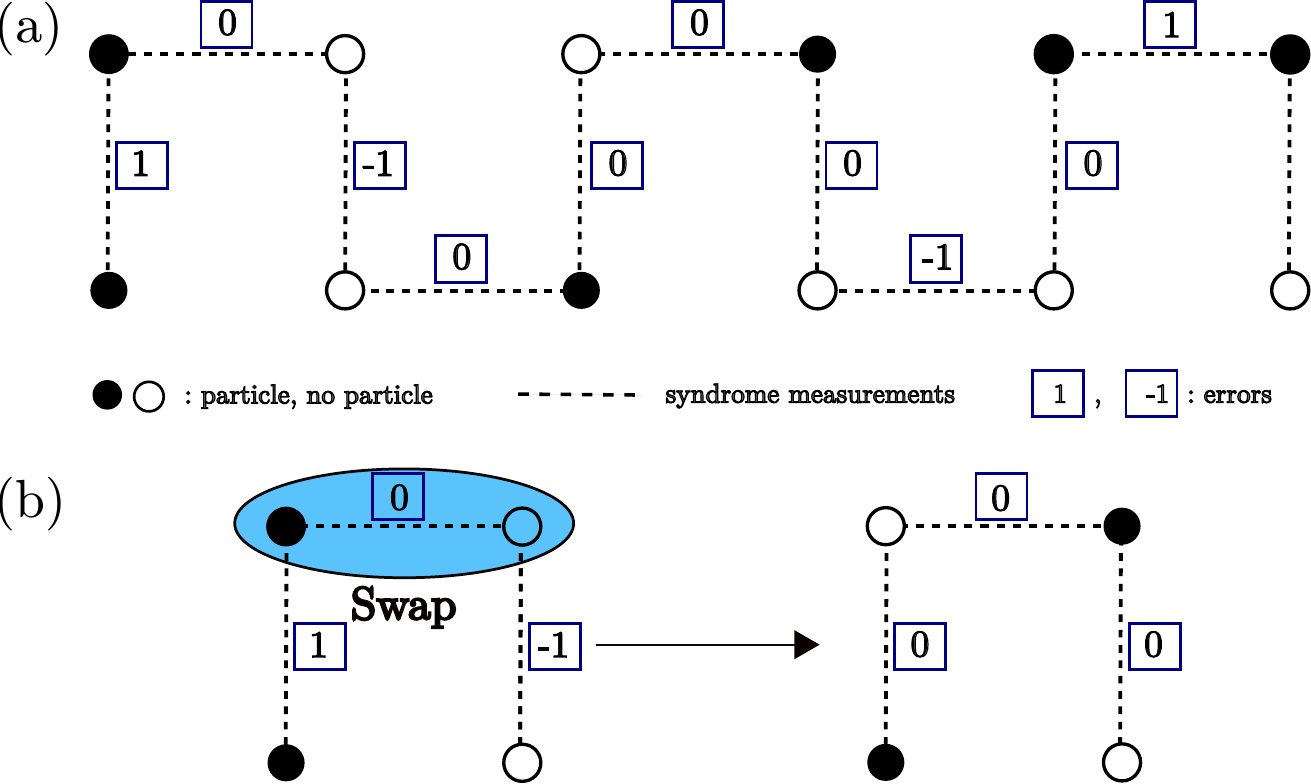}

  \caption{Error correction circuit for density-wave order. (a) The
    local particle configuration is measured and the error syndrome
    for density-wave order is computed along the dashed lines with
    error variables being located on the dual lattice (squares). Two
    neighboring particles signal the presence of a particle error
    (+1), while two empty sites correspond to a hole error (-1). (b)
    Errors are removed by swapping particle positions until a particle
    error is fused with a hole error. All measurement and correction
    operations are carried out without disurbing the spin state.}

    \label{fig:dwerr}
\end{figure}

Let us first describing the error correction algorithm for
density-wave order. As the first step, we identify the position of the
errors by measuring the positions of the particle without disturbing
the spin sector. We perform these measurements on an MPS by applying
our Monte-Carlo algorithm. Errors are identified according to the
measurement results by transversing the system in the pattern shown in
Fig.~\ref{fig:dwerr}, with errors being located on the dual lattice
between the sites of the physical lattice. If there is exactly one
particle on two neighboring site, there is no error (0), while two
particles correspond to a particle error (+1) and two empty sites
correspond to a hole error (-1). Error correction is now being
performed by decorating each error with a walker that explores the
surroundings of the error, with each movement of the walker by one
site corresponds to one timestep of the error correction circuit
\cite{Jamadagni2022}. Once a walker from a particle error encounters a
hole error or vice versa, the errors can be removed by swapping the
particle positions in between such that an error-free configuration is
obtained. Again, this operation is carried out without affecting the
spin state of the particles. Finally, the total number of timesteps
required to remove all errors corresponds to the depth of the
density-wave circuit.

\begin{figure*}[ht]
  \begin{center}
    \begin{tabular}{ccc}
		\includegraphics[width=0.32\textwidth]{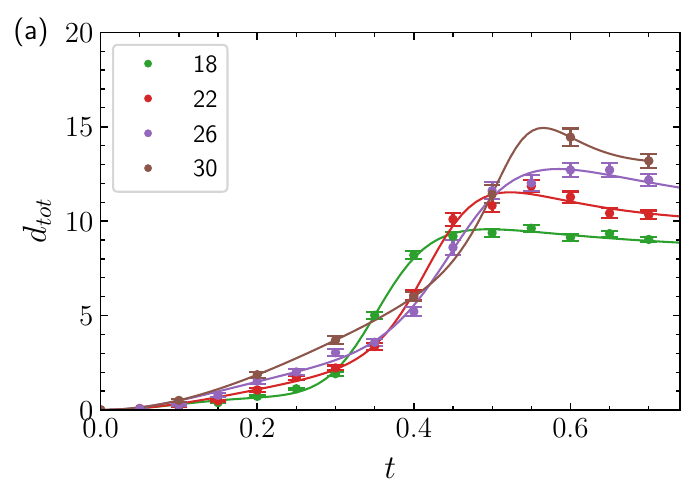}
                &
		\includegraphics[width=0.32\textwidth]{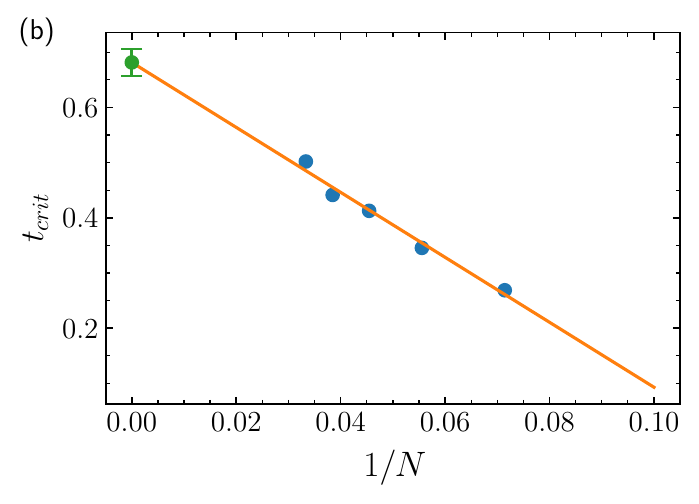}

		\includegraphics[width=0.32\textwidth]{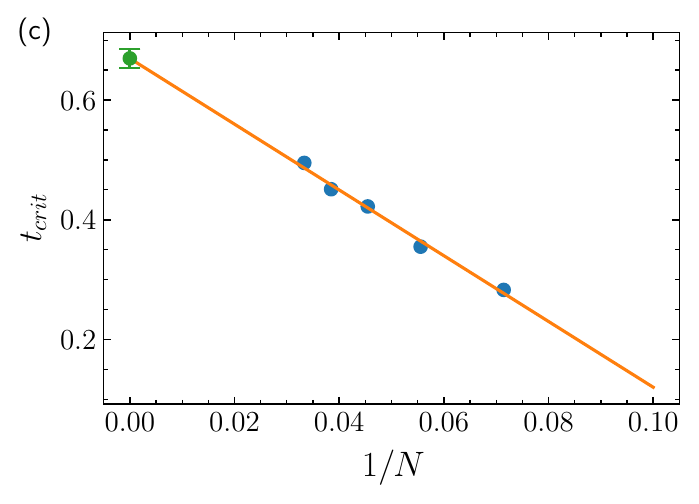}
                \end{tabular}
		\label{critcryst}
        \end{center}
  \caption{Depths of error corrections circuits for the mobile toric
    code. (a) Total circuit depth $d_{tot}$ as a function of the
    hopping strength $t$ for different system sizes. For large $t$,
    the circuit depth diverges in the system size indicating the
    absence of topological order. (b) Finitize size scaling of the
    peak of $\partial d_{tot}/\partial t$, leading to a critical value
    of $t_1=0.67(2)$ in the thermodynamic limit for the breakdown of
    topological order. (c) Correcting only the density-wave order
    leads to a critical value of $t_2 =0.68(2)$ for the disappearance
    of density-wave order. Number of trajectories range from $M=1000$
    for $N=14$ to $M=310$ for $N=30$. }

	\label{fig:data}
\end{figure*}

Once the positions have been corrected to the perfect crystal, we
perform the usual correction of the toric code errors by pairwise
fusion of the $A_v$ and $B_p$ violations \cite{Jamadagni2022}. The
circuit depth of this part is added to the circuit depth of the
density-wave ordering $d_{dw}$ to obtain the total circuit depth
$d_{tot}$ for detecting topological order. Figure \ref{fig:data} shows
the total circuit depth for systems up to $N=30$ particles. MPS
simulations were carried out using the TenPy library
\cite{Hauschild2018} using a maximum bond dimension of $\chi=3000$. We
note that this bond dimension is required to achieve convergence of
the circuit depth in the Luttinger liquid phase. We locate the phase
transition points by looking at the peaks of the derivatives $\partial
d_{tot}/\partial t$ and $\partial d_{dw}/\partial t$,
respectively.
The peaks of the derivatives are obtained by fitting a rational
polynomial function of the form $f(x) = \frac{a x^2 + b x^4 + e x^6}{1
  + c x^2 + d x^4 + f x^6}$ to the simulation data.  Using a
finite-size scaling analysis, we find that topological order
disappears at a critical hopping strength of $t_1 = 0.67(2)$ and
density-wave order disappears at $t_2=0.68(2)$. Hence, our numerical
results strongly suggest that the disappearance of topological order
and density-wave order coincide, pointing to a single simultaneous
phase transition, i.e., there is no intermediate
topologically trivial density-wave ordered phase in
between. See App.~B for additional finite size scaling
  perfomed at a fixed hopping strength.
\begin{figure}[b]
	\centering
	
	\includegraphics[width=0.8\linewidth]{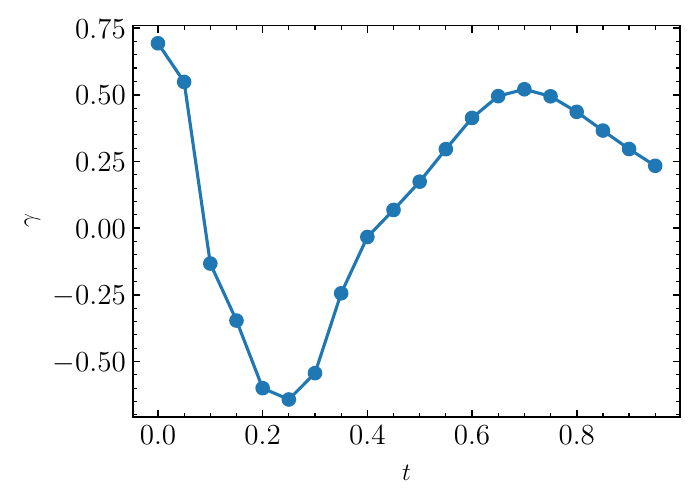}
	\caption{Topological entanglement entropy $\gamma$ of the
          mobile toric code for system sizes up to $N=14$. For $t=0$,
          the TEE corresponds to the toric code value of $\gamma=\log
          2$, but strong finite size effects immediately lead to a
          reduction in the TEE even though the system is still in its
          topologically ordered phase.}
	\label{fig:teefalsch}

\end{figure}

\emph{Comparison to TEE calculations.} We now try to investigate the
topological properties of the mobile toric code using the topological
entanglement entropy. As a first obstacle, we notice that the
exponential scaling limits the accessible system size to at most
$N=14$, which is much smaller than the error correction computations
when using comparable computational resources. For zero hopping, the
TEE is the same as for the toric code on a cylinder, i.e., we have
$\gamma = \log 2$. For large values of $t$, we again expect a
transition to a topologically trivial phase having $\gamma =
0$. However, finitize-size effects are expected to appear as a
rounding of the step-like behavior of the TEE
\cite{Morampudi2014,Jamadagni2018}. Technically, the TEE is computed
by calculating the entanglement entropy for large system sizes and
extrapolating back to $L=0$ to obtain the subleading term according to
Eq.~\ref{eq:tee}. Figure \ref{fig:teefalsch} shows the behavior of the
TEE as a function of the hopping strength $t$. As can be clearly seen,
the results are completely inconclusive once the hopping is switched
on, having little resemblance to the expected step function
behavior. Even for very small hopping strengths, the TEE starts to
deviate from the toric code value of $\gamma = \log 2$, indicating a
loss of topological order much earlier than is actually the
case. These findings unambigously demonstrate the advantages of the
error correction approach to topological order.

\emph{Summary and outlook.---} In summary, we have presented an
efficient numerical strategy to compute topological order in a quantum
many-body system. Our approach relies on an operational definition of
topological order and is exponentially faster than computing the
topological entanglement entropy. We have applied our computational
framework to the example of an extension toric code where particles
are mobile, finding that we can successfully obtain the phase
diagram. Our approach naturally generalizes to higher-dimensional
tensor network states such as projected entangled-pair states
\cite{Orus2014,Cirac2021}, as well as to topological order in
non-equilibrium situations
\cite{Bardyn2013,Yarloo2018,Petiziol2024}. Finally, we would like to
point that the quantities required to compute topological order can be
measured in state-of-the-art quantum simulator experiments
\cite{Semeghini2021,Chen2023}, while experimentally obtaining the
topological entanglement entropy requires quantum state tomography and
is again exponentially hard.

\begin{acknowledgments}
  We thank Roman Orus for fruitful discussions. This work was funded by the Volkswagen Foundation and by the Deutsche Forschungsgemeinschaft (DFG, German Research
Foundation) within Project-ID 274200144 -- SFB 1227 (DQ-mat, Project No. A04).
\end{acknowledgments}

\bibliography{../bib/bib}	
\appendix

\section{Extension to 2D systems}

In one-dimensional systems, it is possible for the
  entanglement entropy to also exhibit a nonvanishing constant term
  $\gamma$ when the system is in a symmetry-protected topological
  (SPT) phase \cite{Zou2016}. When considering a quasi-1D system,
  additional considerations are therefore required to decide whether
  the system will eventually acquire intrinsic topological order when
  the second dimension is extended to infinity.

The concept of the replica correlation length has been introduced to
address this issue. This characteristic length scale influences the
behavior of the subleading term in the entanglement entropy. If the
internal symmetry responsible for the SPT order is absent, the
spurious subleading term $\gamma$ decays exponentially with $L$,
characterized by the replica correlation length.

From numerical simulations of the perturbed toric code
  \cite{Jamadagni2018}, it is known that the correct topological order
  for the 2D model can already be extracted from the corresponding SPT
  phase in a quasi-1D realization. This can be seen from the
  calculation of the modular $S$ matrix, which is characteristic for a
  topologically ordered phase \cite{Zhu2013}, and which for the
  quasi-1D SPT ordered toric code is identical to its 2D
  counterpart. From these considerations, we can conclude that the SPT
  phase of the quasi-1D mobile toric code can be extended to the
  intrinsically topologically ordered phase in 2D.

\section{Finite Size Scaling at fixed hopping strength}

\begin{figure}[h]
	\centering
	\includegraphics[width=0.7\linewidth]{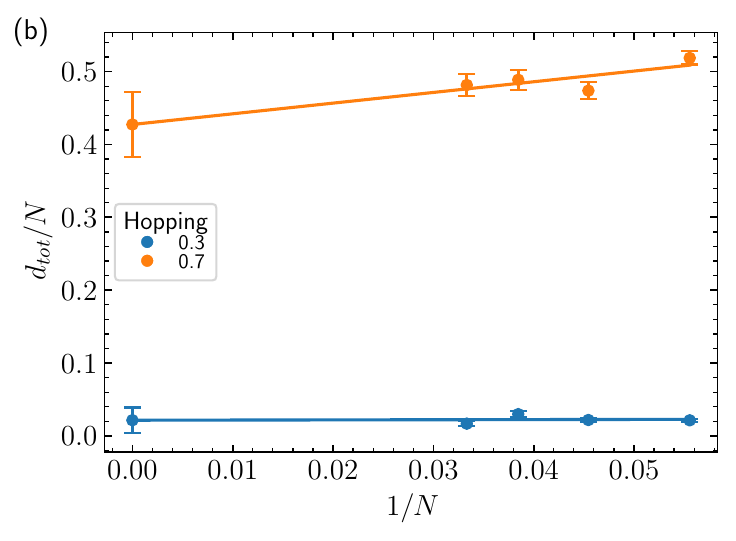}
	\caption{Scaling analysis of the normalized circuit depth
          $d_{tot}/N$ as a function of the inverse system size $1/N$. A
          linear fit is used to extrapolate the behavior in the
          thermodynamic limit.}
	\label{fig:scaling}
\end{figure}

To investigate the scaling properties of correction times for a fixed hopping strength, we analyze their dependence on system size. In Figure \ref{fig:scaling}, we present the circuit depth normalized by the system size as a function of the inverse system size. Normalizing by the system size has the advantage that the quantity becomes zero in the thermodynamic limit for an ordered phase and constant in a disordered phase. Similar to the quantitatively more precise analysis based on the derivative of the circuit depth in the main text,  a clear phase transition can be identified between hopping strengths $t=0.3$ and $t=0.7$.

\clearpage

\end{document}